**Low Thermal Budget High-k/Metal Surface Gate for buried donor-based devices**

Evan M. Anderson, DeAnna M. Campbell, Leon N. Maurer, Andrew D. Baczewski, Michael T. Marshall, Tzu-Ming Lu, Ping Lu, Lisa A. Tracy, Scott W. Schmucker, Daniel R. Ward, Shashank Misra

*Sandia National Laboratories, Albuquerque New Mexico, 87185*

**Abstract**

Atomic precision advanced manufacturing (APAM) offers creation of donor devices in an atomically thin layer doped beyond the solid solubility limit, enabling unique device physics. This presents an opportunity to use APAM as a pathfinding platform to investigate digital electronics at the atomic limit. Scaling to smaller transistors is increasingly difficult and expensive, necessitating the investigation of alternative fabrication paths that extend to the atomic scale. APAM donor devices can be created using a scanning tunneling microscope (STM). However, these devices are not currently compatible with industry standard fabrication processes. There exists a tradeoff between low thermal budget (LT) processes to limit dopant diffusion and high thermal budget (HT) processes to grow defect-free layers of epitaxial Si and gate oxide. To this end, we have developed an LT epitaxial Si cap and LT deposited $Al_2O_3$ gate oxide integrated with an atomically precise single-electron transistor (SET) that we use as an electrometer to characterize the quality of the gate stack. The surface-gated SET exhibits the expected Coulomb blockade behavior. However, the gate's leverage over the SET is limited by defects in the layers above the SET, including interfaces between the Si and oxide, and structural and chemical defects in the Si cap. We propose a more sophisticated gate stack and process flow that is predicted to improve performance in future atomic precision devices.

**1. Introduction**

Exponential increases in tooling costs for successively smaller metal-oxide-semiconductor (MOS) transistor generations are becoming untenable as feature sizes shrink below a linear dimension of 10 nm (~ 30 silicon atoms). To investigate potential device types and architectures before committing large capital investments for manufacturing, smaller scope research devices are necessary. This includes using tools that are inherently limited in scope to study proof-of-concept devices that demonstrate the physical viability of new device technologies beyond the next manufacturing node. One path to developing an understanding of device physics at the absolute limit of atoms themselves is atomic precision advanced manufacturing (APAM) [1]. In APAM, fabrication of nanoscale devices is achieved using a scanning tunneling microscope (STM) to pattern devices on the surface of Si. With this technology devices such as single-electron transistors (SETs) can be produced and used as sensitive electrometers to characterize the impact of different fabrication steps for process development. To date, many atomically precise (AP) devices such as tunnel junctions [2], SETs with a single quantum dot island [3-6], and a SET with a pair of independently controlled quantum dots [1] have been produced using delta-layer doping of P in Si above the solid solubility limit. However, such devices rely on intrinsic Si as a dielectric for in-plane gates rather than MOS surface gates. This limits the applicability of this pathway to the microelectronics industry, which is based on MOS field effect transistors.

The challenge of integrating MOS surface gates on top of AP devices derives from the need to maintain a low thermal budget [1]. There exists a tradeoff between low thermal budget (LT) processes to prevent dopant diffusion and high thermal budget (HT) processes to promote low material defect densities. The APAM process involves patterning a hydrogen mask [7-9] to design an AP device and incorporate a high, non-equilibrium concentration of dopants at the Si surface [10, 11]. Epitaxial Si encapsulation and subsequent process steps must occur at LT to prevent P diffusion and segregation both in the growth direction [12] and laterally. For traditional APAM in-plane gates, the dielectric is intrinsic Si. It is thus limited by the band gap of Si with no offset between the device and the gate. For example, a transistor with an in-plane gate 38 nm from the channel leaks at 0.5 V at 80 mK [13], while a MOS gate with an oxide dielectric barrier of 20 nm or thinner would be expected to sustain 10 V with minimal leakage [14]. Minimizing leakage and increasing transconductance are critical for the low-power operation of digital logic devices, further motivating a move from in-plane to surface gates. Finally, operation at room temperature or higher is critical to effectively investigate the path forward for improving the performance of complementary metal-oxide-semiconductor (CMOS) devices. To date APAM devices have only operated at cryogenic temperatures (e.g., 4K), "freezing out" lightly doped Si and rendering it an effective gate dielectric. For room temperature operation this will not be the case, necessitating an insulator between the gate and the APAM channel.

Standard thermal $SiO_2$ provides the simplest path to a clean dielectric but it is not thermally compatible with APAM. The required processing temperature will cause aggressive diffusion of the dopants. A MOS-gated SET has been created once through a unique *in situ* Si and O codeposition at room temperature [15] to grow $SiO_2$ after Si encapsulation of an APAM SET also including an in-plane gate. This additional surface gate was demonstrated to improve the performance of the in-plane gate [13]. An alternative is to use an *ex situ* deposited oxide, which allows investigation of high-k dielectrics that decrease tunneling leakage and maintain high capacitance for effective gating with thinner layers. Deposited high-k dielectrics have already been implemented in industrial fabs, underscoring their compatibility with CMOS and the feasibility of integrating APAM devices with CMOS process flows.

In this paper we discuss the performance of an APAM SET integrated with a high-k gate oxide, $Al_2O_3$, deposited by atomic layer deposition (ALD). We use this SET as an in-channel electrometer to evaluate the quality of the gate stack and to understand device performance. We characterize this gate stack with cross-sectional scanning transmission electron microscopy (STEM) and secondary ion mass spectrometry (SIMS) to identify process defects and discuss the impact of these defects on SET performance. Finally, we propose a gate material stack that promises to overcome these process defects to improve future device performance.

**2. Experimental Procedure**

Front-end-of-line (FEOL) fabrication, STM device patterning, and back-end-of-line (BEOL) processing were executed using previously described methods [1, 16]. FEOL produced chips with As implants for contacts to the APAM devices and alignment marks that are discernable in the STM. Next, oxide removal, device patterning via STM hydrogen depassivation lithography (HDL), $PH_3$ dosing, and Si molecular beam epitaxy (using an MBE Komponenten Si sublimation (SUSI) source) occurred in ultra-high vacuum (UHV) chambers to produce an APAM SET encapsulated with 30 nm of Si. An STM image of this SET is shown in figure 1(a), where the bright regions are the patterned areas that have had the H removed to allow the incorporation of P that comprises the nanostructures. BEOL consisted of etching vias and depositing Al to contact the As implants that contact the APAM devices. A surface gate was added to the APAM SET by first depositing the gate dielectric, annealing in forming gas in a rapid thermal annealer, and finally depositing an Al gate electrode. During process development we compared $SiO_2$ and $Al_2O_3$ films deposited using plasma-enhanced chemical vapor deposition and ALD, respectively (see supplementary material). The gate dielectric of this SET is 30 nm of $Al_2O_3$ deposited by ALD at 200 °C with trimethyl aluminum and water precursors. A cross-sectional schematic of a completed device is shown in figure 1(b). It is important to note that unlike a traditional field effect transistor, the epitaxial Si immediately beneath the gate oxide is effectively part of the gate stack. The channel is the P-doped APAM SET. Finally, electrical testing of the fabricated SET was conducted at 4K.

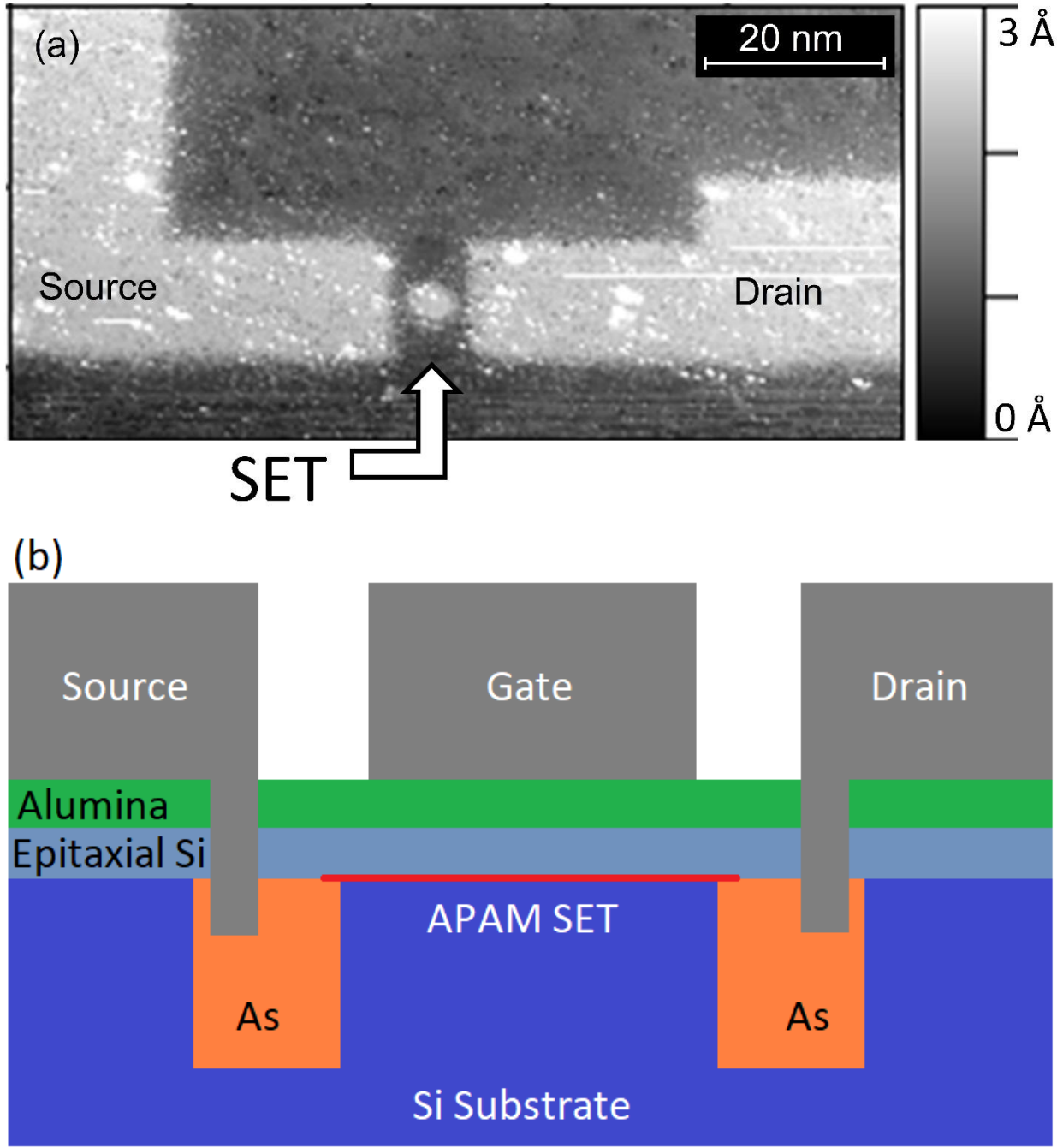

Figure. 1. (a) STM image of SET after H depassivation lithography before creating the MOS gate stack. (b) Schematic cross section of a completed MOS gate stack consisting of $Al_2O_3$ for the oxide and Al for the electrode on top of an APAM device and As implants to contact the APAM SET.

## 3. Results and Discussion

We infer that the SET survived the BEOL processing and gate deposition by the presence of the characteristic diamond structure in the source-drain current, as a function of source-drain and gate bias, that typifies Coulomb blockade [17] (figure 2(a)). The presence of Coulomb diamonds is evidence that the P donors remained in place with minimal diffusion. If the dopants diffuse a few nanometers in-plane, they will fill the tunnel barriers between the island and the leads (figure 1(a)) and the device will behave as a nanowire. If the dopants diffuse significantly either in-plane or out-of-plane, the carrier concentration will become too low to conduct at 4K and the device will behave as an open circuit. Here we evaluate the performance of the top-gated device in terms of the leverage of the gate on the SET, which we define in terms of the ratio of the charging energy of the SET (diamond amplitude) to the voltage applied to the gate needed to change the charge state of the SET by one electron (diamond period). For this top-gated device the island is a single 5-7 nm radius feature (figure 1(a)), with uncertainty as to the final atomic and effective electronic device dimensions after fabrication. We observe a charging energy of approximately 6 meV for approximately 0.3 V applied on the gate between charging events, indicating that the gate leverages the SET by ~20 meV/V. Overall, our data compare well both to a previous effort with a leverage of ~33 meV/V for an in-plane gated APAM SET that also included an unconventional top gate [13], and our own data for an in-plane gated APAM SET produced without a top gate (see supplementary material). Assuming ideal dielectric permitivities for both our 30 nm Si cap and 30 nm $Al_2O_3$ dielectric and an island radius of 6 nm, we calculate an island-gate capacitance of approximately 1 aF and an island self-capacitance of approximately 5 aF. This yields a capacitance-defined leverage of 0.2, compared to 0.36 for an in-plane gate in prior work [13].

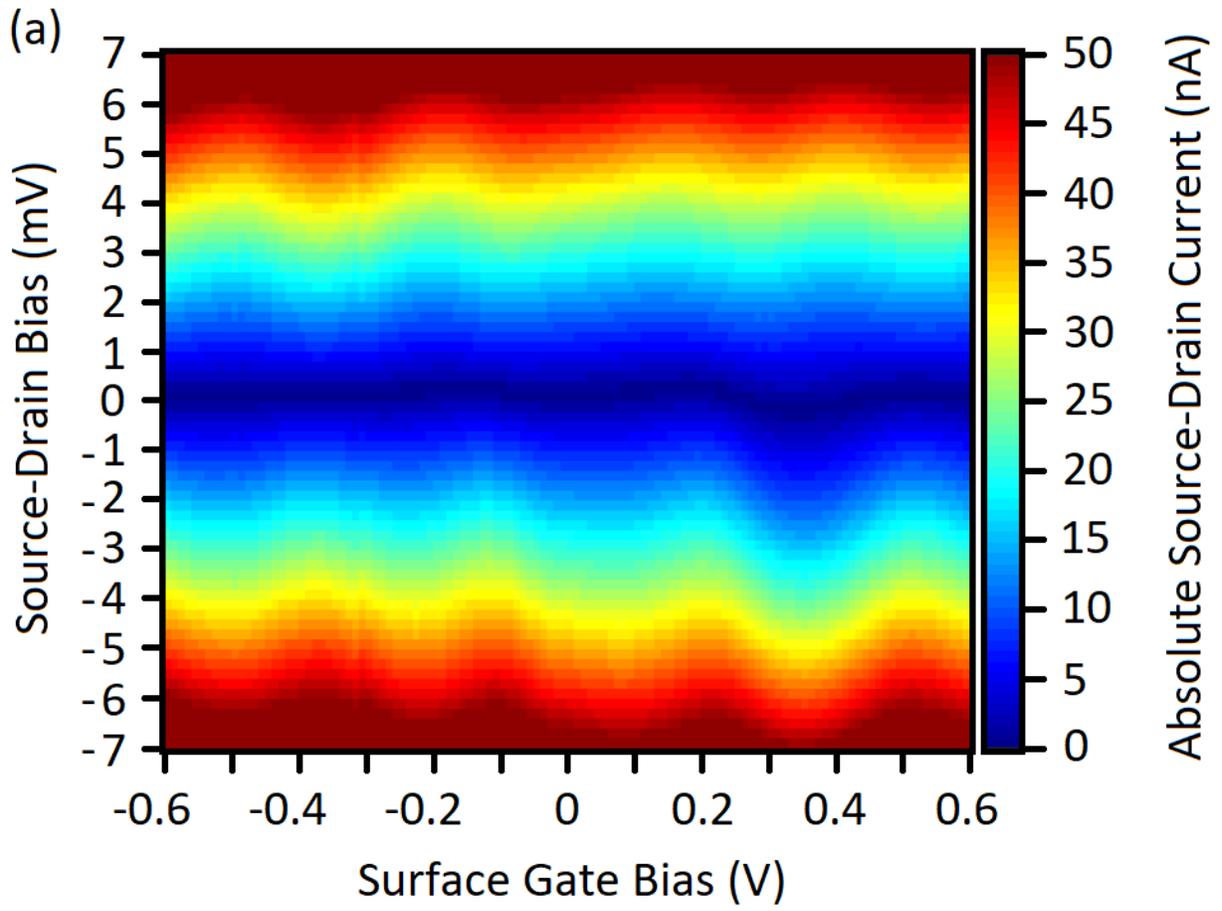

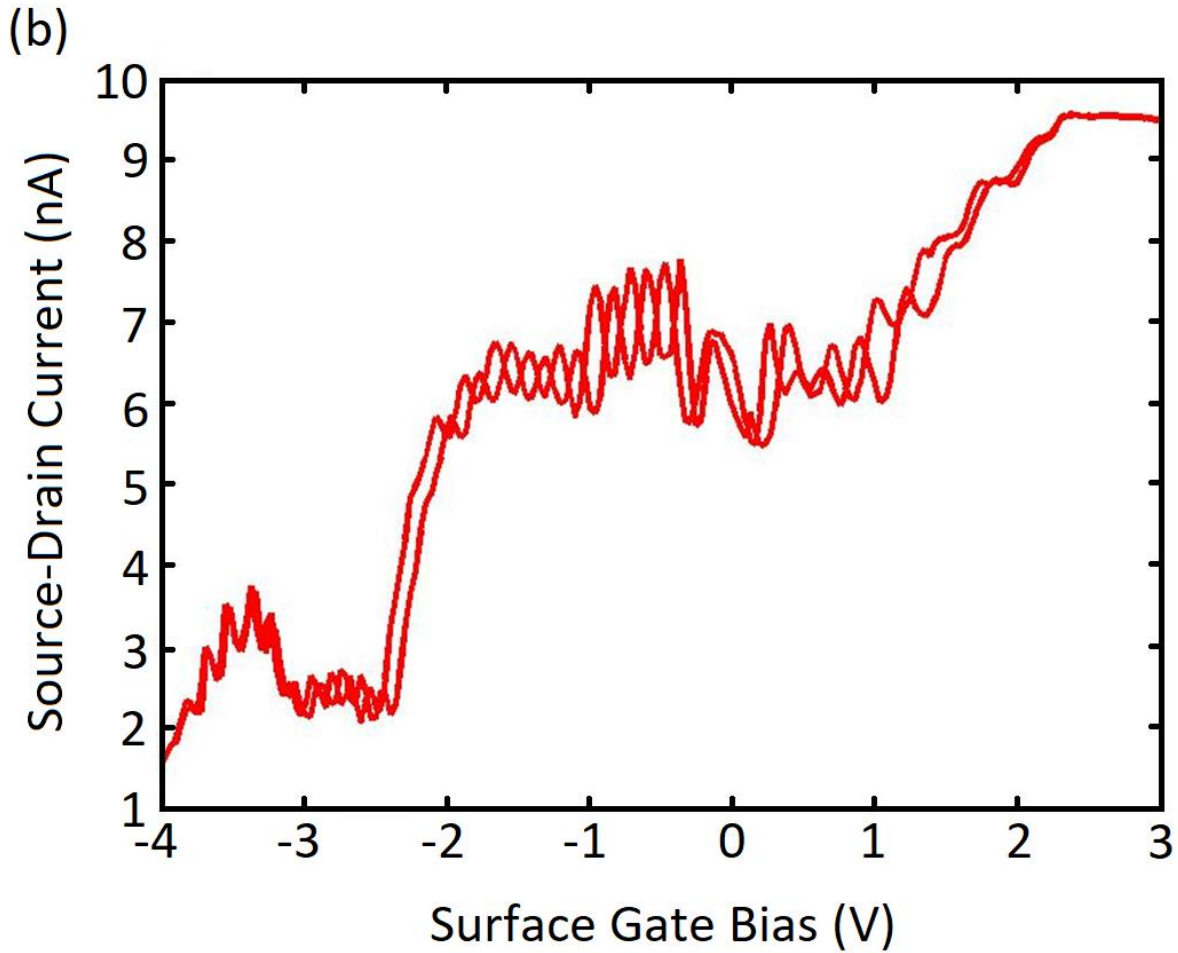

Figure 2(a). Absolute source-drain current as a function of gate bias and source-drain bias for an APAM SET with a top gate exhibiting the characteristic diamond structure of Coulomb blockade, leading to ~6 meV charging energy (diamond amplitude) for a change of ~0.3 V on the gate (diamond period). (b) Source-drain current for the SET for forward and reverse sweeps of the gate bias at a source-drain bias of 1 mV.

The results above demonstrate that a low-temperature ALD oxide is a viable path forward for producing an APAM device with a MOS surface gate. The ability to apply a higher gate bias with a MOS surface gate (see supplementary material) is valuable for quantum information applications, where a large gate bias is required to control the number of electrons on the island. For digital logic applications, high transconductance translates to low operating voltages and power consumption. However, our surface-gated SET has not achieved superior performance; it has similar leverage to another SET with an APAM in-plane gate. Numerical modeling suggests qualitatively that we should examine the quality of the $Al_2O_3$ gate oxide and interfaces between materials to understand how to optimize the gate stack (see supplementary material). This is reinforced by the data in Figure 2(b), showing that the SET source-drain current (at a source-drain bias of 1 mV) suffers from hysteresis on subsequent gate bias sweeps in opposite directions over the range of -4 V to 3 V. This hysteresis suggests that there exist trap states in the epitaxial Si, the $Al_2O_3$, the $Si/Al_2O_3$ interface [18], and/or the $Al/Al_2O_3$ interface. These traps could be caused by structural and chemical defects, which would be consistent with LT Si homoepitaxy [19] (though this can be mitigated with extremely slow Si growth [20]) and LT ALD of $Al_2O_3$ on Si [21], respectively.

Material characterization reveals defects and impurities that are consistent with the observed gate hysteresis. Impurities in the silicon cap could create trap states that cause hysteresis or generate carriers that screen the field applied by the gate, noting that the epitaxial Si is effectively part of the gate stack above the P-doped APAM channel. Figure 3 shows a SIMS depth profile of an epitaxial cap, grown under similar conditions to that of the SET, with incorporated O and Al. These concentrations are $7\times10^{19}$ atoms/cm$^3$ O and $4\times10^{18}$ atoms/cm$^3$ Al in the cap with $1\times10^{17}$ atoms/cm$^3$ O and $1\times10^{15}$ atoms/cm$^3$ Al (roughly the detection limit for Al in Si) in the substrate. Similar results have been reported previously for this LT Si epitaxy process on another similar sample, though that sample exhibited a higher concentration of chemical defects (C, O, and N) at the cap/substrate interface [22]. The C, N, and O concentrations are influenced by the Si growth temperatures [22] and are limited by the chamber background pressure during growth, which is approximately $1\times10^{-8}$ Torr. We suspect that the Al impurities come from the SUSI source, as similarly prepared films capped with a different silicon source in the same chamber have not contained Al. While O is typically a deep level state, it can exist at a shallow energy level on its own [23], or in a complex with N [24], and become electronically active. Al is a shallow acceptor and impacts gate performance.

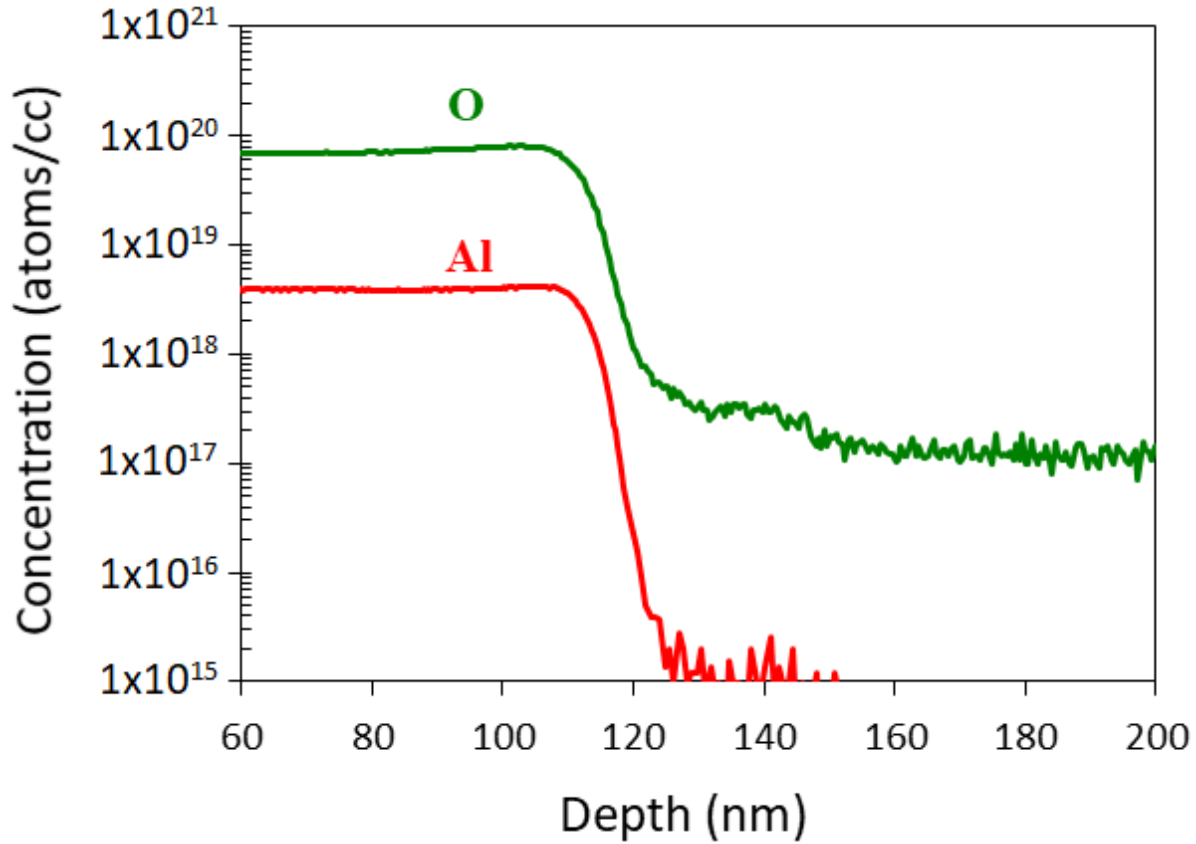

Figure 3. SIMS depth profile of epitaxial Si grown under similar conditions to that of the APAM SET and APAM nanowire.

STEM reveals further material defects in the gate stack. For an APAM nanowire with ALD $Al_2O_3$ on top of an epitaxial Si cap deposited similarly to the SET, STEM reveals an uncontrolled $SiO_2$ layer approximately 4 nm thick at the interface between the $Al_2O_3$ and the Si, which is visible as the darker region between the lighter $Al_2O_3$ and lighter crystalline Si (figure 4). We expect a high interfacial trap density due to the poor interface between $Al_2O_3$ and Si in our samples [25]. The presence of this $SiO_2$ layer effectively decreases the average dielectric constant of the gate stack and, in turn, the gate's capacitance over the SET. Additionally, figure 4 shows that there are structural defects in the epitaxial Si near the oxide, which are visible as darker regions in the high angle annular dark field (HAADF) image partially obscuring the bright, periodic columns of Si atoms. These defects might indicate partial amorphization of the cap or the formation of (111) twin domains near the Si/oxide interface.

Amorphous Si or twin boundaries could introduce electronically active states and contribute to the observed gate hysteresis (figure 2(b)).

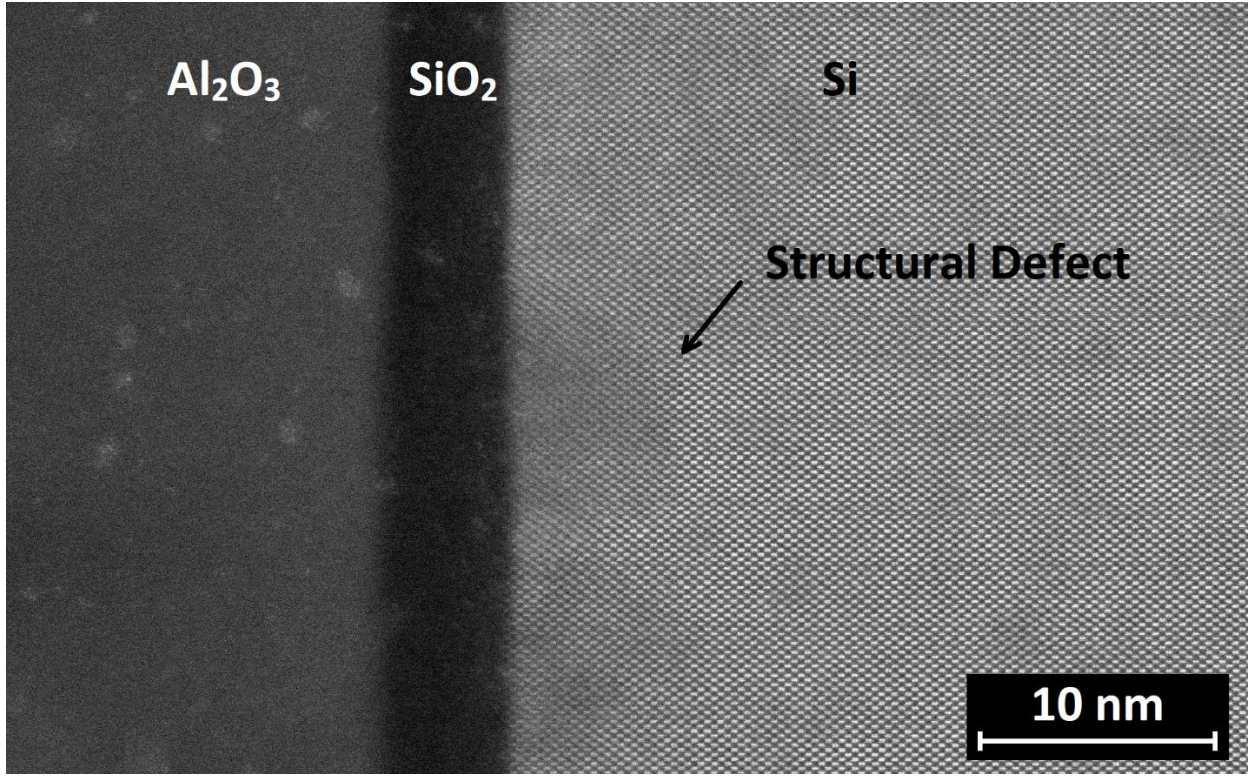

Figure 4. STEM HAADF image of cross sections of a nanowire with ALD Al$_2$O$_3$ atop the epitaxial Si, prepared under similar conditions to the SET. The bright spots in the oxide are due to damage from focused ion beam used to prepare the STEM sample. Note: the nanowire is not discernable in this cross section and might not be contained within it.

In general, the interface with the highest density of interface states should have the largest impact on device performance in terms of gate leverage. Qualitatively, in a simple linear model, the Si/Al$_2$O$_3$ interface is likely the most important. Since the gate leverage over the SET depends on both the island self-capacitance and the gate-island capacitance, the leverage will be decreased by electronically active defects at all of the interfaces in the gate stack. The island self-capacitance will only be influenced by the defects in the island, that is, over a small region of the Si/P-doped Si interface. The contribution of defects elsewhere in that plane to the self-capacitance is negligible. The decrease in leverage will largely be driven by the decreased gate-island capacitance. Within a particular range of bias conditions, the gate-island capacitance will be reduced relative to what it would be in the absence of interface traps by an amount determined by the density of trap states, within those bias conditions, at all interfaces between the gate and the island. The gate-island capacitance is effectively reduced by the leverage that the gate has over the traps between the gate and island, so reducing the density of trap states is critical to realizing ideal performance. However, the density of trap states in a given device is dependent on the processing conditions and the quality of the interfaces, which is a subject of further study for the material stack and processing workflow at hand. Finally, the $Al/Al_2O_3$ interface is less important than the $Si/Al_2O_3$ interface. Traps at these interfaces must be compensated by either the channel or the gate. The further traps are from either interface, the stronger the field and the effect on leverage will be. The $Si/Al_2O_3$ interface is far from either the channel or the gate and will have a larger effect on leverage, than the $Al/Al_2O_3$ interface. Moreover, as a good metal, Al will be more effective than Si in screening the charged trap states, reducing the field that they produce at the island.

Many of the materials defects examined above can be reduced, or their effects mitigated, by adding to the complexity of the relatively simple material stack presented here. Our observed $SiO_2$ interface layer is consistent with, though thicker than, previous reports [14, 21, 26]. This could be mitigated by intentionally depositing a thinner $SiO_2$ [27] or $Si_3N_4$ [28] layer to act as a diffusion barrier before $Al_2O_3$ deposition. Additionally, the electrode/oxide interface can be improved by depositing barrier layers like TiN or TaN to prevent the Al from reducing the oxide when in direct contact [29, 30]. An additional TiN or TaN layer on top of Al would protect the Al from formation of a native oxide and improve electrical contact to the gate itself. Furthermore, deposited $HfO_2$ could replace $Al_2O_3$ as it has a higher dielectric constant, allowing a smaller "effective oxide thickness" [25].

The performance of the SET can be further improved through optimizing the geometry of the gate stack and the STM-patterned part of device. Assuming the island self-capacitance is only dependent on the in-plane geometry, gate leverage can be increased by making the layers of the gate stack thinner.

Thinner layers move the gate closer to the SET (device channel), thus increasing the capacitance and therefore the capacitance-defined leverage of the gate over the SET. The epitaxial Si cap could be thinned to approximately 5 nm before surface scattering and the effects of limited out-of-plane P diffusion on device resistivity become significant [31]. An ideal oxide layer, assuming no pinhole defects, could also be thinned to a few nanometers before leakage becomes a concern [25]. Taking the more conservative thickness of 10 nm for both the Si cap and $Al_2O_3$, would make the gate stack a third of the thickness of the current device and triple the gate capacitance and leverage, assuming a simplistic model of the electrostatics. Finally, the geometry of the SET island and leads, including their sizes, shapes, and spacing could be further optimized to increase gate leverage, confinement energy, number of electrons on the island, and period of the Coulomb diamonds [5, 6, 32].

## 4. Conclusions

We have developed a simple, low thermal budget high-k/metal gate stack, using ALD $Al_2O_3$ as the high-k dielectric, that is compatible with APAM devices to enable investigation of physical principles for future digital device generations. We have evaluated the performance of this surface-gate stack for an APAM SET and observed the characteristic Coulomb blockade with a gate leverage of 20 meV/V, indicating that the SET survived the BEOL processing to deposit the gate stack. However, this gate is hysteretic, which is the consequence of process defects including low-quality interfaces and structural defects in the Si cap as observed by STEM, in addition to O and Al that were unintentionally incorporated into the Si cap as observed by SIMS. To improve the performance of the surface gate, we propose a new gate stack with a diffusion barrier such as TiN or TaN between the Al and high-k dielectric layers, a controlled deposition of $SiO_2$ or $Si_3N_4$ to act as a diffusion barrier for the interface between the Si cap and high-k dielectric, investigation of alternative dielectrics, and improvement of the Si epitaxy. A more advanced material stack, combined with additional optimization of the APAM process, should reduce the hysteresis, and improve the leverage of the gate. By reducing the role of impurities and defects, a more advanced material stack may also improve the ability of simple capacitive models to capture the measured behavior. Optimizing the Si/Al$_2$O$_3$ interface is likely to have the largest impact on improving gate leverage. Additionally, thinning the epitaxial Si and dielectric to 10 nm each should triple the leverage of the gate over the SET.

The surface gate stack demonstrated here, along with our proposed improvements, advance APAM as a technology for investigating physical principles that will aid the development of next-generation transistor nodes and Si:P-based quantum devices. Both applications require high-performance gates. Digital logic devices require high transconductance to switch devices with low power consumption. This requires gates that are as close to the device channel as possible with minimal leakage, both of which are enabled by high-k gate dielectrics. Digital logic devices must also be operated at room temperature or higher to be practical. APAM in-plane gates are prone to leakage, which is exacerbated at room temperature, suggesting that surface gates are a necessary advance for exploring this application space. To access a wide range of charge configurations for donor-based quantum devices, the large energy scales for donors drives the need for higher transconductance devices and larger voltages of operation. Leakage through the intrinsic Si dielectric imposes a lower bound on the separation between the gate and device, effectively upper bounding the achievable capacitance. In contrast, a surface gate with a high-k dielectric can be placed much closer to the channel and can be driven to higher bias without concerns about leakage.

**Acknowledgement**

The Far-reaching Applications, Implications and Realization of Digital Electronics at the Atomic Limit (FAIR DEAL) project is supported by the Laboratory Directed Research and Development program at Sandia National Laboratories, and was performed, in part, at the Center for Integrated Nanotechnologies, a U.S. DOE, Office of Basic Energy Sciences user facility. Sandia National Laboratories is a multimission laboratory managed and operated by National Technology and Engineering Solutions

**Supplementary Material: Low Thermal Budget High-k/Metal Surface Gate for buried donor-based devices**

Evan M. Anderson, DeAnna M. Campbell, Leon N. Maurer, Andrew D. Baczewski, Michael T. Marshall, Tzu-Ming Lu, Ping Lu, Lisa A. Tracy, Scott W. Schmucker, Daniel R. Ward, Shashank Misra

*Sandia National Laboratories, Albuquerque New Mexico, 87185*

Additional atomic precision advanced manufacturing (APAM) devices were produced and additional gate stack process development was conducted beyond what was described in the main text of the manuscript. As in the manuscript, front-end-of-line processing produced scanning tunneling microscope (STM) compatible chips as substrates for APAM device processing using previously described methods [1, 16]. This includes several 2-terminal STM-patterned, $PH_3$-dosed nanowires, a tunnel junction, and a single electron transistor (SET) with a single in-plane gate and an island comprised of two quantum dots in parallel (figure S1). To evaluate dielectrics for a metal-oxide-semiconductor (MOS) surface gate on atomically precise Si:P devices, we evaluated two low-temperature deposited oxides: $SiO_2$ deposited by plasma-enhanced chemical vapor deposition (PECVD) at 250°C, and $Al_3O_3$ deposited by atomic layer deposition (ALD) at 200°C with trimethyl aluminum and water as precursors. Low deposition temperatures were used to avoid diffusion of the P atoms from the atomically precise device channels. Oxide deposition was followed by a forming gas anneal at 300°C for 15 min. Electrical conduction through non-gated, 40 nm wide nanowires was measured at 4K to determine resilience of patterned devices to dielectric deposition and annealing (figure S2). The nanowire with the $SiO_2$ dielectric exhibited non-Ohmic conduction (figure S2(a)), while the nanowire with the $Al_2O_3$ dielectric (figure S2(b)) behaved similarly to nanowires with no dielectric, exhibiting Ohmic conduction [16]. Consequently, the top-gated SET discussed in the main text was fabricated with ALD $Al_2O_3$.

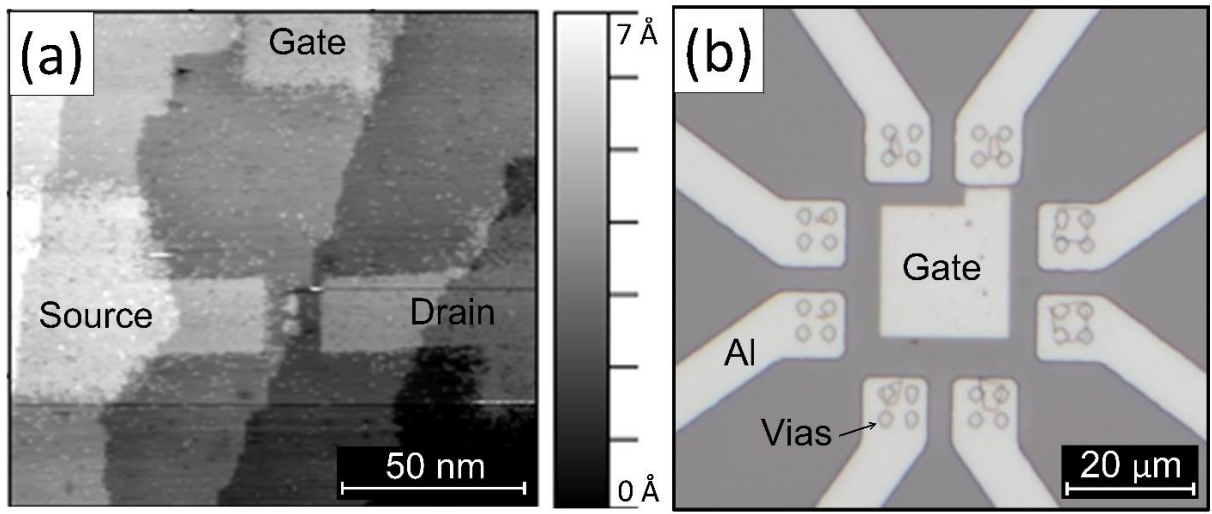

Figure S1. (a). STM image of a SET with an in-plane gate after STM patterning. (b). Optical micrograph of a complete APAM SET with a MOS top gate

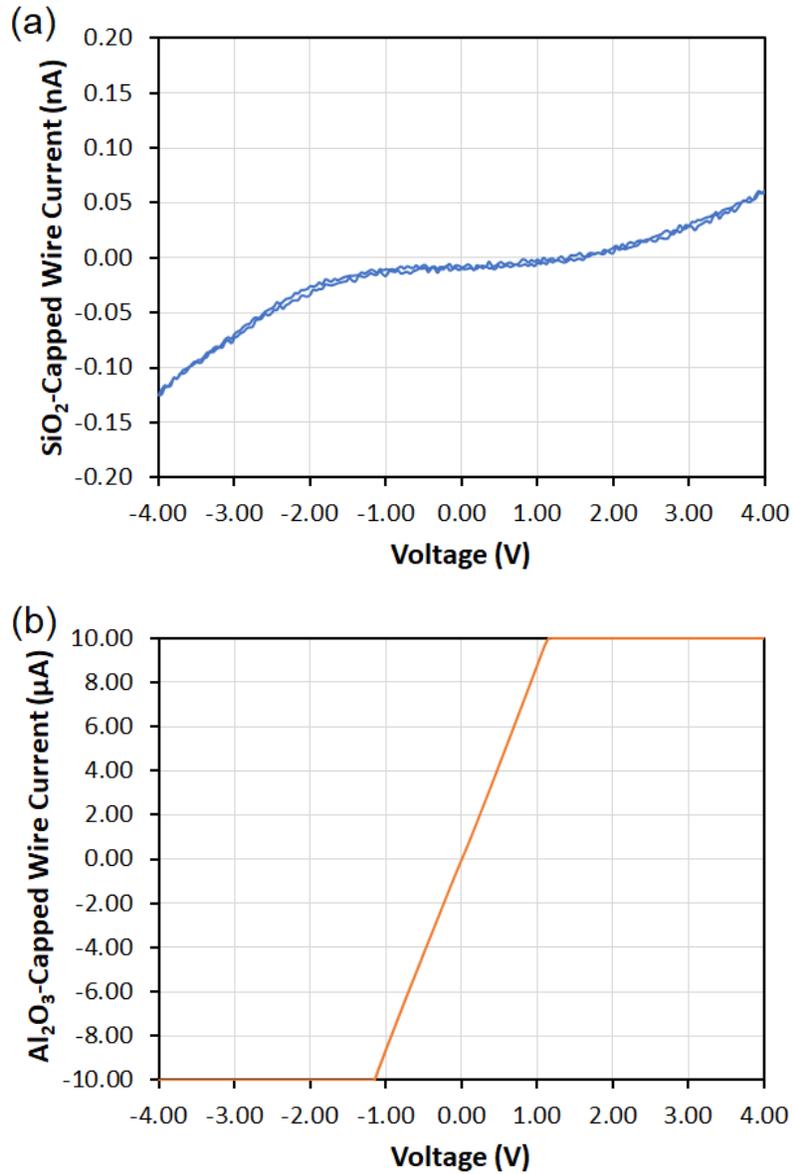

Figure S2. (a). I-V curve with current in nanoamps for a P nanowire capped with 30 nm of epitaxial Si and 50 nm PECVD SiO$_2$ b. I-V curve with current in microamps for a P nanowire capped with 30 nm of epitaxial Si and 30 nm ALD Al$_2$O$_3$.

As alluded to in the main text, here we compare the performance of the top- and in-plane-gated devices in terms of the leverage of the gate on the SET. The current through the SET as a function of the source-drain and gate bias voltages is shown for the MOS-SET in figure 2(a) of the main text and for the in-plane gated SET here in figure S3. Both sets of data are plotted on the same 0 to 50 nA scale for ease of comparison. The characteristic diamond structure that typifies Coulomb blockade [17] is evident for both. Here we define leverage in terms of the ratio of the charging energy of the SET to the voltage applied to the top- or in-plane gate needed to change the charge state of the SET by one electron. In comparing the two devices, we are careful to note that they differ not only in the relative orientation of the gate but in the dimensions of their SET islands. For the top-gated device the island is a single 5-7 nm feature, while it is a pair of 2-3 nm features separated by 6 nm for the in-plane-gated device (see figure 1(a) in the manuscript and figure S1(a) here). For the top-gated SET we observe a charging energy of approximately 6 meV for approximately 0.3 V applied on the gate between charging events, indicating that the gate leverages the SET by ~20 meV/V. For the in-plane-gated SET a charging energy of 15 meV is evident for approximately 0.4 V applied to the gate between charging events, indicating a leverage of approximately 38 meV/V. In both devices, the presence of Coulomb diamonds is evidence that the P donors remained in place with minimal diffusion after back-end-of-line processing and deposition of the gate oxide. Overall, our data compare well to a previous effort where an APAM SET had been demonstrated with an in-plane gate and a MOS top gate with an unconventional physical vapor deposition of $SiO_2$ [13]. Nevertheless, in comparing the leverage of the two devices it is surprising to note that they are comparable, though with the in-plane SET performing slightly better, regardless of the orientation of the gate.

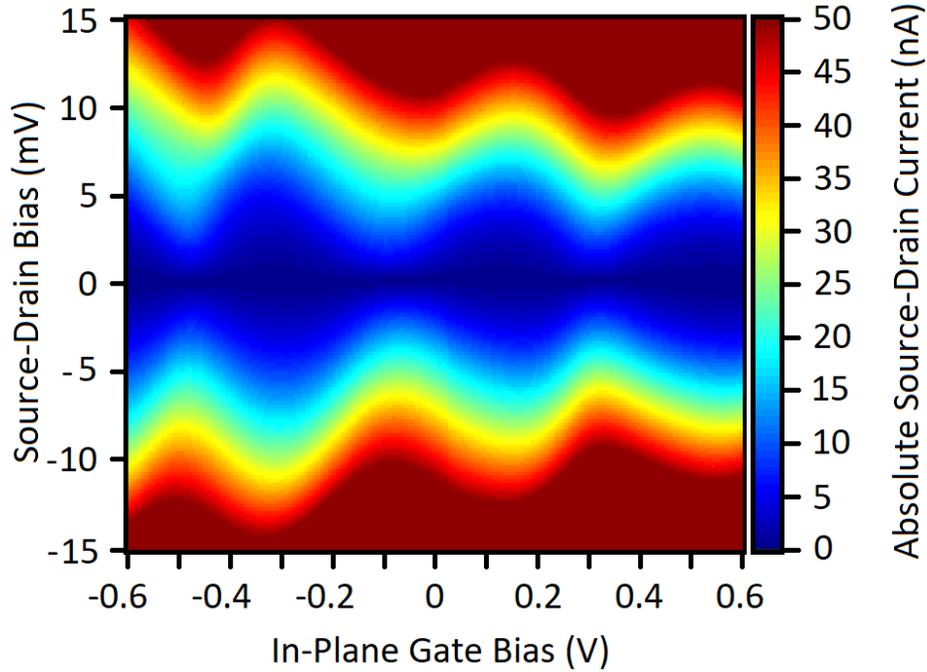

Figure. S3. Absolute source-drain current as a function of gate bias and source-drain bias for an APAM SET with an in-plane gate, leading to ~15 meV charging energy for ~0.4 V on the gate.

Modeling can be used to check the assumption that the surface gate should out-perform the in-plane gate while accounting for some expected non-idealities. Both devices are operating far from the single- to few-electron regimes based on the short period and number of Coulomb oscillations. Consequently, the SETs' gating behavior can be rationalized in terms of a classical electrostatic model in which all P components are treated as perfect conductors with no internal quantum degrees of freedom. We also assume ideally sharp interfaces between materials. COMSOL Multiphysics was used to solve the associated Poisson equation and to extract the SET island's self-capacitance, as well as its mutual capacitance to the gate and lead electrodes. These are plotted in figure S4(a) for the top gated SET and figure S4(b) for the in-plane gated SET for a range of plausible island radii (R) and relative permittivities ($\varepsilon_{cap}$). For each SET, we investigated a range of R due to uncertainty of the final geometry of the island, given the stochastic nature of the dopants and the possibility of dopant diffusion. We investigated a range of values of $\varepsilon_{cap}$ due to its dependence on the defect density in the capping silicon [33-35], starting from the ideal value of $\varepsilon_{cap}=\varepsilon_{Si}=11.7$ and increasing it by up to a multiplicative factor of 4. For all values of R and $\varepsilon_{cap}$ the island's self-capacitance dominates the mutual capacitances by a factor greater than 2. This suggests that each SET's performance is dominated by the self-capacitance of the metallic island.

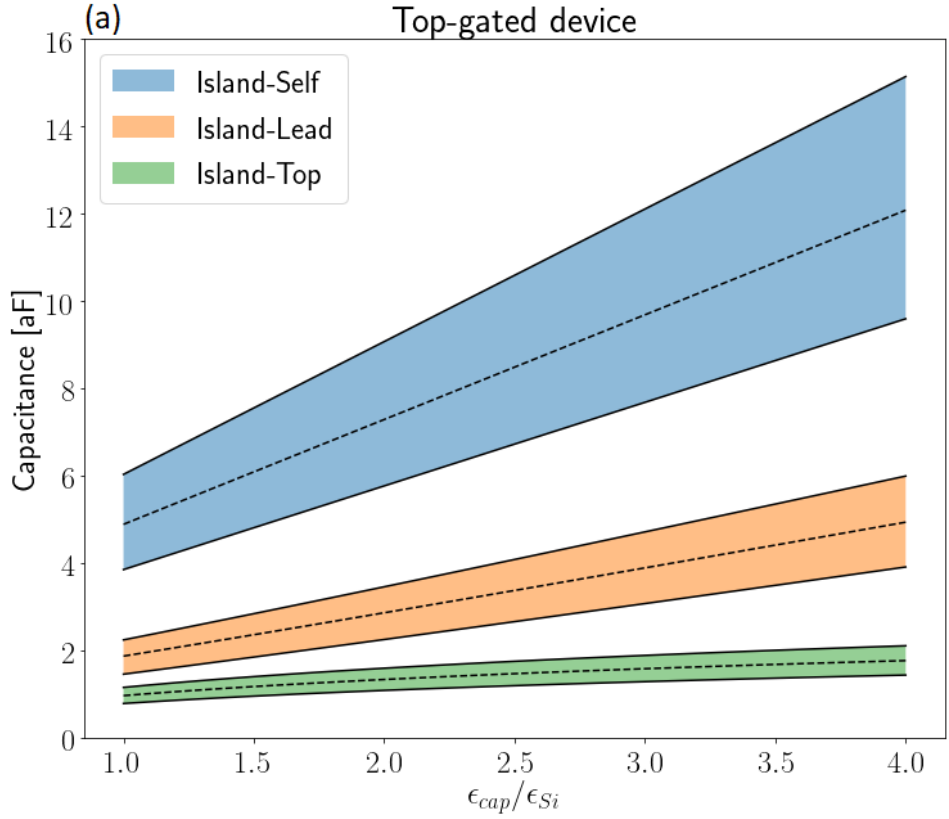
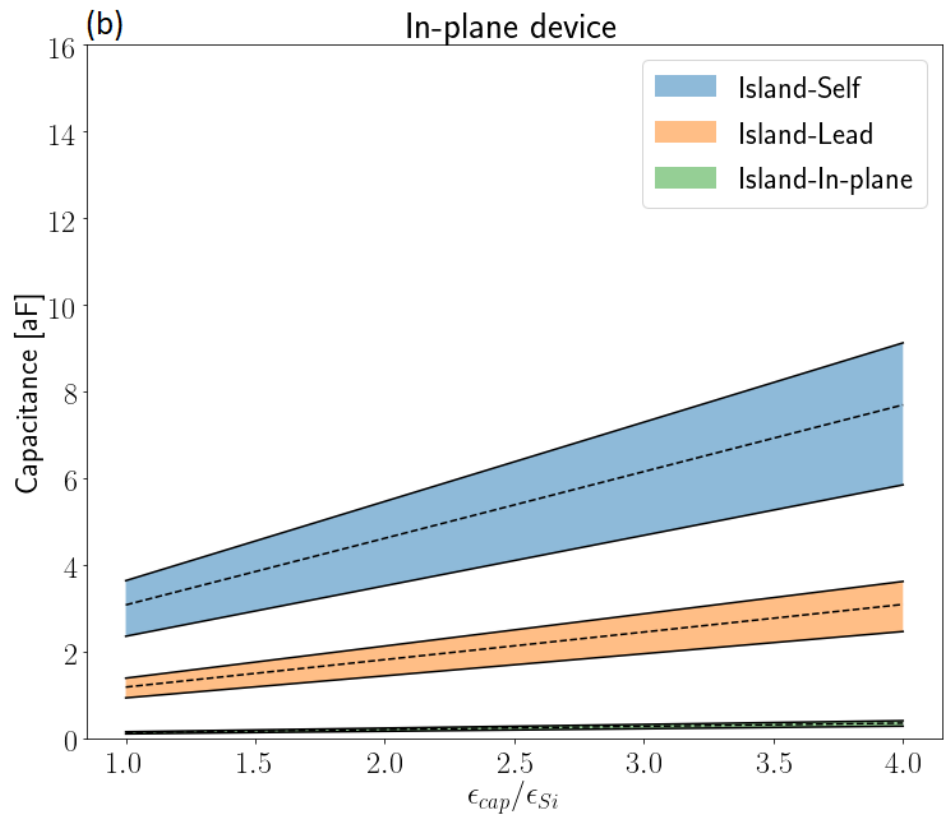

Figure S4. Device capacitances from an electrostatic model of the devices as a function of the dielectric constant of the capping layer and the SET island dimensions for (a) the top-gated device and (b) the in-plane-gated device. In all cases, the spread in values for different SET island radii are indicated by the shaded regions. For the top-gated device with a single island these vary from 5 to 7 nm (6 nm dashed). For the in-plane-gated device with two islands, these vary from 2 to 3 nm (2.5 nm dashed) for a fixed inter-island separation of 6 nm.

To determine whether the surface gate should out-perform the in-plane gate when including uncertainties in geometry and dielectric constant, we compare the leverages of the top- and in-plane gates on the SET islands in Figure S5. Here the leverages are defined as the ratio of the respective island-gate capacitance to the self-capacitance. For a pristine capping layer without defects, we expect that the top-gated device will have four times the leverage of the in-plane gated device. Even accounting for uncertainty in the dielectric constant and device geometry, the simple model suggests our top-gated SET should perform better than our in-plane gated SET, while the experimental data indicates the opposite. Digging deeper, our results show that the top-gated device's leverage is sensitive to the dielectric properties of the silicon cap in a way that the in-plated gated device's is not. While increasing the capping layer's dielectric constant due to the presence of defects reduces the leverage for the top-gated device, the in-plane gated device is insensitive to this change due to the difference in the geometries. The fact that the measured leverages of the two devices are comparable can then be rationalized in terms of the differential impact of defects in the capping layer on the two device designs. Because the in-plane gate and island are coplanar by definition, the change in the dielectric constant will impact both commensurately. In contrast, the island-self capacitance increases more rapidly as a function of the capping dielectric constant than the island-gate capacitance, leading to a decrease in leverage with a degradation in the quality of the capping layer. More generally, complications in the material stack mostly affect the leverage of the surface gate, which strongly motivates the materials optimization of the gate stack discussed in the main text.

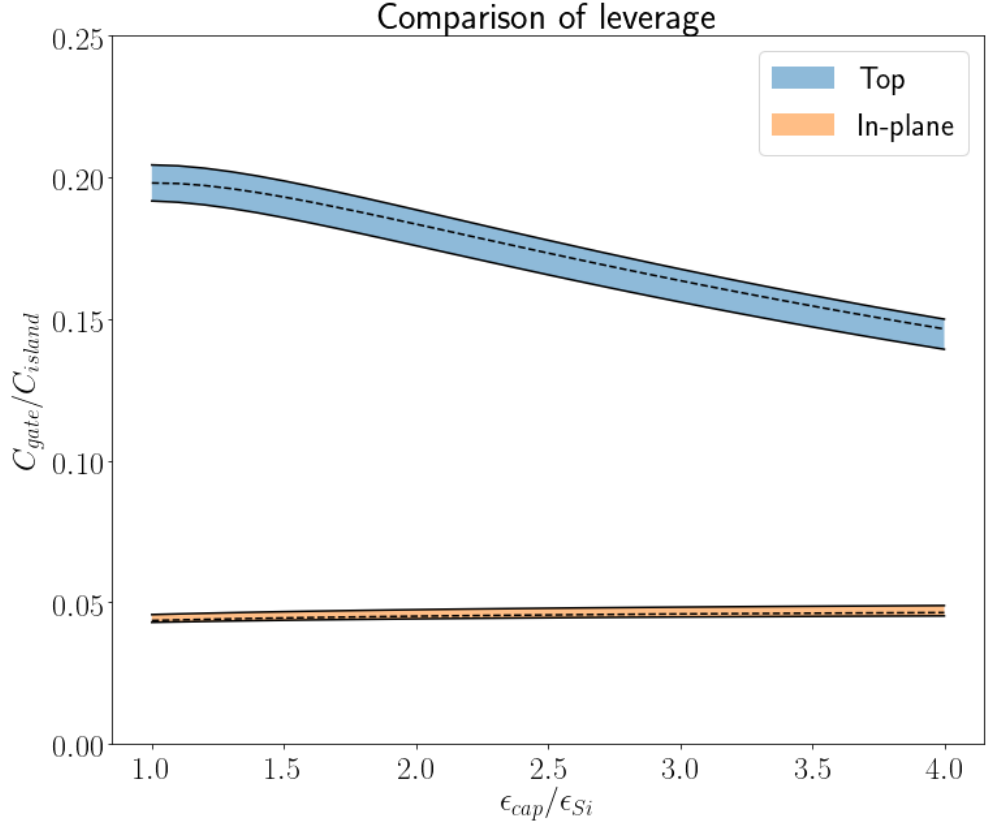

Figure S5. A comparison of the leverage of the top and in-plane gated devices, defined in terms of the ratio of the island-gate and island-self capacitances, as a function of the dielectric constant of the capping layer.

Despite issues with the leverage of the surface gate, we observe that the MOS gate breaks down at higher voltage than the in-plane gate does, making it useful for certain applications, like gating quantum dots, despite having a comparable leverage. Figure S6 shows leakage current for the surface-gated SET from the main text and the in-plane-gated SET. The in-plane gate is approximately 55 nm from the nearest island in its SET (50 nm from the edge of the leads) and the surface gate is approximately 60 nm from its SET. The voltages at which leakage occurs for the in-plane gated SET are typical, about +/- 1 V, while the surface-gated SET does not show comparable leakage until past +/- 3V (figure S6(b)). However, the breakdown voltage of the MOS gate is not ideal, though this discrepancy can be explained by our device layout. Because the gate's Al contact is shorted to one of the As-implanted contacts (the gate is connected to the vias in figure S2(b), shown schematically in figure S7), we are effectively measuring the breakdown across the p-n junction formed between the As implants and the p-type substrate instead of leakage through the gate oxide. This is indicated by the diode-like curve showing what appears to be rectification. This can easily be remedied in future devices by updating the device layout and masks to allow a dedicated contact for the surface gate that does not short to an As-implanted region. We expect this improvement to allow applying +/- 10 V gate bias with minimal leakage, as would be expected for such a thick dielectric layer [14].

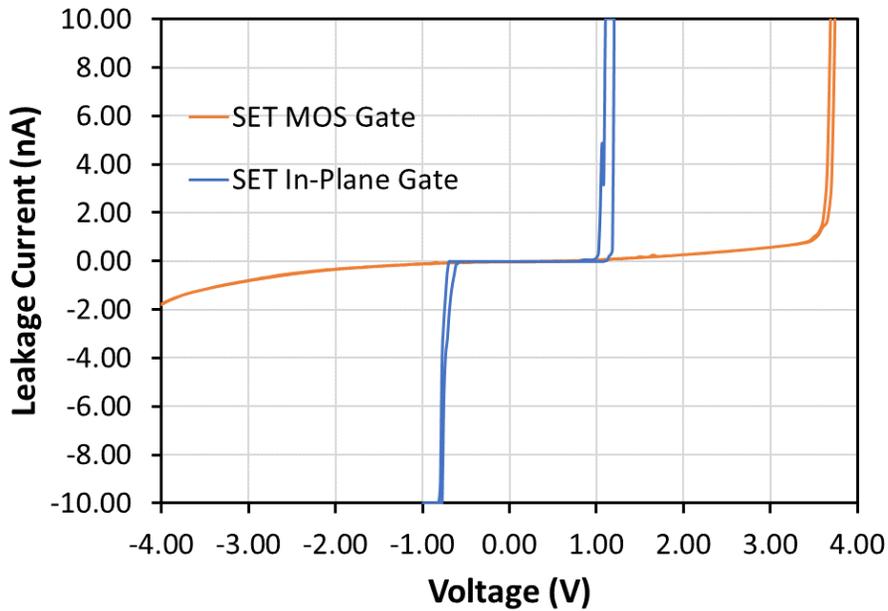

Figure S6. Plots of leakage current vs. applied voltage for an APAM SET with a MOS (Al/Al$_2$O$_3$/Si) surface gate (the device in the main text) and an APAM SET with an APAM Si:P/Si/Si:P in-plane gate.

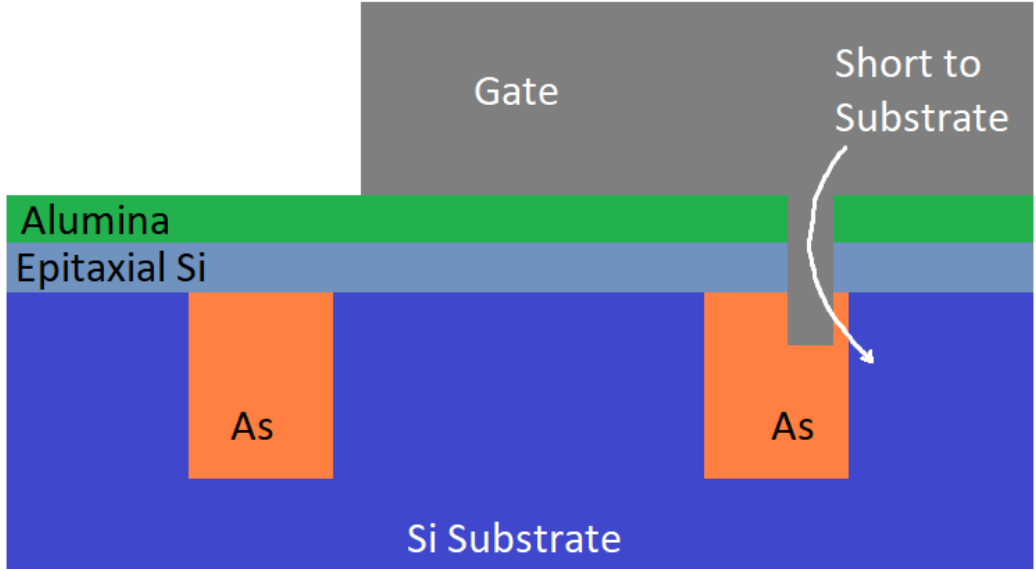

Figure S7. Cross sectional schematic of the surface illustrating a leakage path from the gate, through the As implant, to the p-type substrate.

In summary, we have discussed process development of our MOS surface gate stack and compared the performance of APAM devices with surface gates and in-plane gates. An APAM nanowire with a CVD $SiO_2$ layer on top was unexpectedly resistive, while an APAM nanowire with an ALD $Al_2O_3$ layer on top exhibited the expected ohmic conduction at 4K. Thus, all subsequent MOS gates in this study used the ALD $Al_2O_3$ as the gate oxide. We observe that our surface gated MOS SET performs well compared to an in-plane gated SET, with both devices exhibiting Coulomb blockade behavior. Capacitance modeling of the devices was used to qualitatively understand the behavior of both devices. Additionally, all devices with a MOS gate sustained higher applied bias before leaking compared to in-plane gated devices, as expected.

**Acknowledgement**

The Far-reaching Applications, Implications and Realization of Digital Electronics at the Atomic Limit (FAIR DEAL) project is supported by the Laboratory Directed Research and Development program at Sandia National Laboratories, and was performed, in part, at the Center for Integrated Nanotechnologies, a U.S. DOE, Office of Basic Energy Sciences user facility. Sandia National Laboratories is a multimission laboratory managed and operated by National Technology and Engineering Solutions of Sandia, LLC., a wholly owned subsidiary of Honeywell International, Inc., for the U.S. Department of Energy's National Nuclear Security Administration under contract DE-NA-0003525. This paper describes objective technical results and analysis. Any subjective views or opinions that might be expressed in the paper do not necessarily represent the views of the U.S. Department of Energy or the United States Government.